\documentclass[
reprint,
 aps,
 superscriptaddress,
bibnotes,
]{revtex4-2}

\usepackage{physics}
\usepackage{dsfont}
\usepackage{color,newfloat}
\usepackage{graphicx}
\usepackage{dcolumn}
\usepackage{bm}
\usepackage{array}
\usepackage{amsmath,amssymb,amsthm}
\usepackage{amsfonts}
\usepackage{bbold}
\usepackage{mathtools}
\usepackage{multirow}
\usepackage{hhline}
\usepackage[dvipsnames]{xcolor}
\usepackage{hyperref}
\usepackage[T1]{fontenc}
\hypersetup{
    colorlinks=true,
    linkcolor=Maroon,
    citecolor=Maroon,
    urlcolor=Maroon
}
\usepackage[caption = false]{subfig}
\usepackage{lineno}

\usepackage{makecell}

\begin{document}


\title{
Long-distance quantum communication using concatenated ring graph codes
}

\author{Love Pettersson}
\email{love.pettersson@nbi.ku.dk}
\affiliation{Center for Hybrid Quantum Networks (Hy-Q), Niels Bohr Institute, University of Copenhagen, Blegdamsvej 17, 2100 Copenhagen, Denmark.}

\author{Anders S. Sørensen}
\email{anders.sorensen@nbi.ku.dk}
\affiliation{Center for Hybrid Quantum Networks (Hy-Q), Niels Bohr Institute, University of Copenhagen, Blegdamsvej 17, 2100 Copenhagen, Denmark.}


\begin{abstract}
To realize long-distance quantum communication, it is crucial to design quantum repeater architectures that can deal with transmission losses and operational errors.
Code concatenation of photonic graph codes is a promising way to achieve this; however, existing concatenated codes that can correct both transmission losses and operational errors are extremely hardware-demanding. 
We propose a one-way quantum repeater architecture based on concatenated ring graph codes and linear optical Bell-state measurements.
We construct a scheme to generate the concatenated ring graph codes using quantum emitters, where the number of matter qubits scales linearly with concatenation depth.
Furthermore, we devise a measurement strategy at each repeater station with a simple experimental setup where photons are measured in the order that they are created and show that entanglement swapping is fault-tolerant to both transmission losses and operational errors.
This allows for long-distance quantum communication ($> 10^4$ km) at a kHZ rate even in the presence of single qubit error rates $\epsilon > 10^{-3}$.
\end{abstract}
\maketitle
\section{Introduction}
To realize large-scale quantum networks, we need to be able to transmit quantum information over long distances.
This is challenging as the transmission channels are noisy and lossy, which degrades the rate and fidelity of the transmitted quantum information~\cite{Kimble2008, QuantumInternetWehner}.
Quantum repeaters overcome this challenge by introducing intermediate stations which splits the transmission length into shorter segments~\cite{BriegelQR}. 
For standard fiber-based transmission channels, transmission decreases exponentially with the length of the channel, whereas the resources for quantum repeaters only scale polynomially.
This makes battling transmission losses more manageable.
There are two main families of quantum repeater protocols~\cite{LukinRepeaterMethodsComparison}.
The conventional two-way repeater protocol relies on heralded entanglement generation between stations through probabilistic photon gates~\cite{LukinTwoWay, NicolasTwoWay, EwertTwoWay}.
This requires long-lived quantum memories, a demanding experimental task, and classical two-way signalling between the chain of repeater stations, which slows down the communication rate.
To overcome the challenges of long-lived quantum memories and the need for two-way signalling, an alternative quantum repeater approach has been proposed, known as one-way repeater~\cite{AllPhotonicRepeaters, SophiaRGS, BorregaardOneWay, BorregardFaultTolerantOneWay, EwertQPCRepeaterFusionGen, LukinQPCRepeater100EmitterGenerated, LeeQPCRepeaterFusionGenerated, HassanRepeaterFancyErrorCorrection}, which is the focus of this work.
This approach redundantly encodes the quantum information in quantum error correction codes to battle transmission losses and operational errors between repeater nodes. 
This allows the entanglement swapping probability between stations to be close to unity, eliminating the need for quantum memories and classical two-way signalling. 
Although the one-way repeaters are promising for high-speed communication, they have their own experimental challenges.
The challenges depend on the requirements of the protocol, which can be divided into two classes; the first class aims to correct for both operational and loss errors~\cite{EwertQPCRepeaterFusionGen, LeeQPCRepeaterFusionGenerated, 
LukinQPCRepeater100EmitterGenerated, EwertQPCRepetearLong, BorregardFaultTolerantOneWay} while the second class aims to only correct for loss errors~\cite{BorregaardOneWay, EconomouRGSGen, SophiaRGS, Hilaire2021RGS, AllPhotonicRepeaters}.
The main challenge for the first class is the required resources at each station to generate the error-correcting codes, which either demand a double-digit or higher number of matter qubits per station or a daunting number of probabilistic entangling gates. 
Instead, the main challenge for the second class is that they cease to function as operational errors increase~\cite{BorregaardOneWay, SophiaRGS, Hilaire2021RGS}.
Furthermore, some of them require photon delay lines to correct for photon emission ordering, which requires low-loss delay lines - an experimentally demanding task. \\
Here, we propose a one-way quantum repeater protocol based on concatenated ring graph codes~\cite{Tom, Codeconcat} and probabilistic two-qubit gates, so-called Bell state or fusion measurements~\cite{browne2005, fusionrules, Gimeno2016}.
The protocol is realized by generating two concatenated rings at each repeater station and performing \textit{logical fusions} for entanglement swapping, similar to the proposals in Refs.~\cite{AllPhotonicRepeaters, Tom}.
We devise a generation scheme for the concatenated rings using a single quantum emitter coupled to memory spin qubits, where the number of memory spin qubits scales linearly with the code depth.
This generation scheme is flexible and can be realized with many different types of quantum emitters, such as atoms~\cite{Yang2022, Thomas2022, thomas2024fusion}, color centers~\cite{vasconcelos2020}, quantum dots~\cite{Schwartz2016, Istrati2020, Appel2022, meng2023deterministic, Coste2023, quandelaQDGraphstates}, or ions~\cite{NetworkTrappedIonsMemory, MonroeIonSpinPhotonInterface}.
With the simple generation scheme, we devise a straightforward measurement procedure for a logical fusion between two concatenated rings where photonic qubits are measured in the order in which they are created, removing the need for delay lines compared to other proposals~\cite{SophiaRGS, BorregaardOneWay, Hilaire2021RGS}.  
Furthermore, we show that this measurement procedure allows for a fault-tolerant logical fusion robust against both photon loss and operational errors.
With these features, we demonstrate that long-distance ($10^4$ km) high-rate quantum communication is possible in the presence of single qubit error rates $\epsilon > 10^{-3}$ using only a single-digit number of matter qubits per station.

\section{Graph states and graph codes}
\label{sec:graphcodes}
The procedure we consider can best be understood in terms of graph states.
Graph states are entangled states that can be represented by a graph where vertices represent qubits and edges represent control-Z entangling gates~\cite{hein, hein2006graphstates}.
More formally, a graph $G(E, V)$ represents a graph state
\begin{equation}
    \label{eq:graph_states}
    \ket{G} = \prod_{i, j \in E}CZ_{i,j}\ket{+}^{|V|}
\end{equation}
where $\ket{+} = \frac{1}{\sqrt{2}} (\ket{0} + \ket{1})$ is a qubit in the plus eigenstate of the Pauli-X operator, and $CZ_{i, j}$ is a control-Z gate between qubits $i$ and $j$.
Graph states belong to the larger class of \textit{stabilizer states}~\cite{gottesman}, where the stabilizer generators of a graph state are $S_i = X_i \prod_{j \in \mathcal{N}(i)}Z_j$, with $i \in V$, $\mathcal{N}(i)$ is the neighborhood of qubit $i$, and $X = \ket{+}\bra{+} - \ket{-}\bra{-}$ and $Z = \ket{0}\bra{0} - \ket{1}\bra{1}$ are the Pauli matrices~\cite{hein2006graphstates}.
These generators generate the full stabilizer group $\mathcal{S} = \langle S_k \rangle_{k=1}^{N}$, where the graph state is an eigenstate of all stabilizer with a $+1$ eigenvalue.
A $N$-qubit graph state can be turned into a $N-1$-qubit graph code encoding one logical qubit initialized in the logical state $\ket{\overline{\psi}} = \frac{1}{\sqrt{2}} (\alpha \ket{\overline{0}} + \beta \ket{\overline{1}})$, where $\ket{\overline{0}}$ and $\ket{\overline{1}}$ represents the eigenstates of logical $Z$.
This is realized by choosing an encoding qubit $M$ and measuring it in the $\{\ket{\psi^*}, \ket{\psi^{\perp*}} \}$ basis~\cite{Tom} obtaining an outcome $\ket{\psi^*}$~\footnote{In the case of $\ket{\psi^{\perp*}}$ outcome, $\ket{\overline{\psi}^{\perp}}$ is retrived instead which can be rotated to $\ket{\overline{\psi}}$ by applying $\overline{X}\overline{Z}$~\cite{Tom}}, where $\ket{\psi^*} = \frac{1}{\sqrt{2}} (\alpha^* \ket{0} + \beta^* \ket{1}) $.
The logical qubit encoded in the graph code can be described by its logical operators, which are $\overline{X} = \prod_{k \in \mathcal{N}(M)}Z_k$, and $\overline{Z} = S_{b}$ for any choice of qubit in the neighborhood of $M$, i.e., $b \in \mathcal{N}(M)$ \footnote{Logical Y is given by the product of logical X and Z $\overline{Y} = i\overline{X}\overline{Z}$.}.
Furthermore, the stabilizer generators of the logical qubit are retrieved from the $N$-qubit graph state as $\langle S_{b}S_l, S_k \rangle_{l \in \mathcal{N}_M \setminus b, k \notin \mathcal{N}(M)}$.
That is, all stabilizer generators with a Pauli-Z support on the encoding qubit are multiplied with $\overline{Z}$ of the graph code.
Since the product of a logical operator and a stabilizer is still a valid logical operator~\cite{Tom, SurfaceCodeLoss}, the complete set of logical operators is $\mathcal{L} = (\overline{X}, \overline{Z}, \overline{Y}) \cdot \mathcal{S}$.

\section{Generation of rings}
\begin{figure}
\includegraphics[scale=0.35]{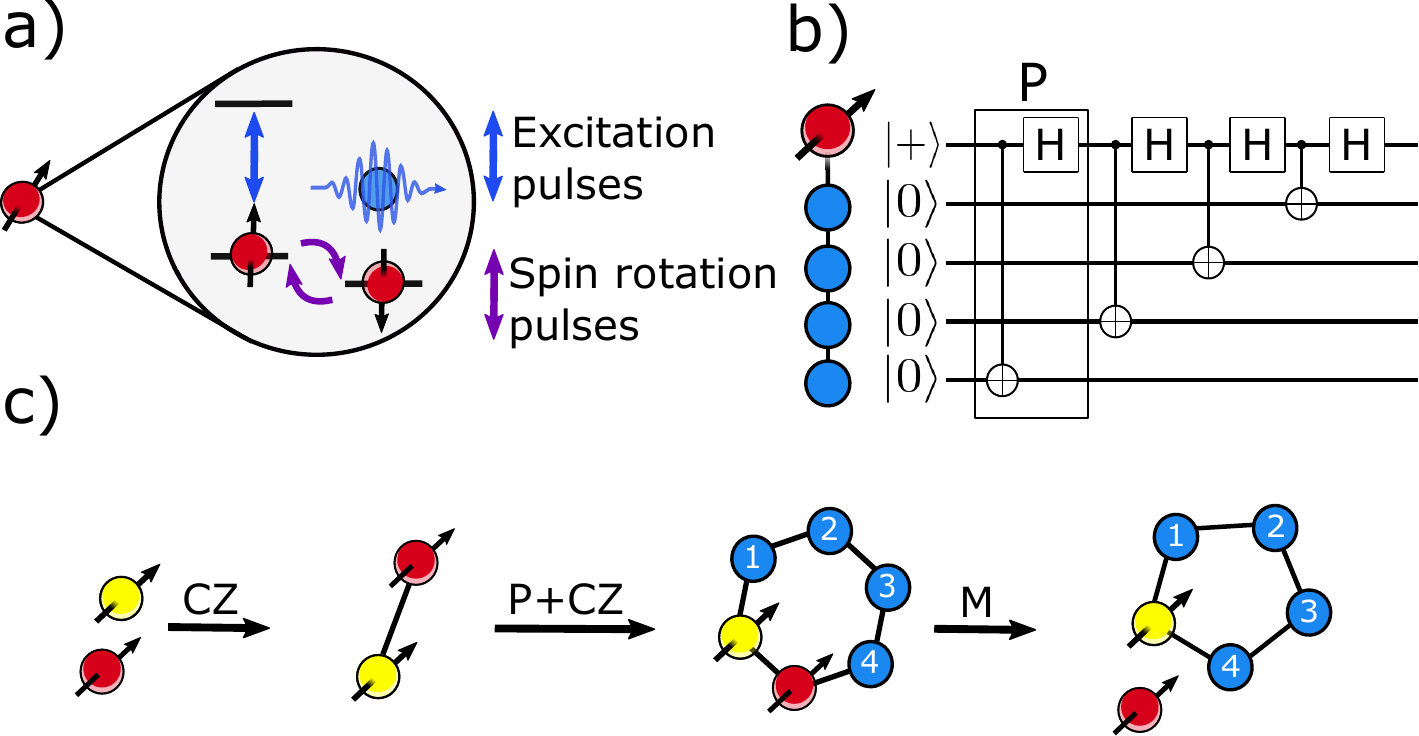}
\caption{a) Schematics of a quantum emitter where photonic qubits are generated by alternating sequences of excitation pulses (blue) and spin rotation pulses (purple). The operation following a suitable pulse sequence can be translated into a graph state picture corresponding to replacing the spin node with a photon node and connecting the spin node to it by a line. This process is referred to as $P$ in the figure, and repeating this process produces a line graph state, as shown in b). c) Generation of a four-qubit ring with one memory spin qubit (yellow) and one optically active spin (red). The generation proceeds by entangling the two spins with a control-Z gate (CZ), then emitting four photonic qubits in a line structure, followed by another entangling gate between the spins and a measurement (M) of the optically active spin.  \label{fig:small_gen_fig}}
\end{figure}
\begin{figure*}[t]
    \centering
    \includegraphics[width=1.0\textwidth]{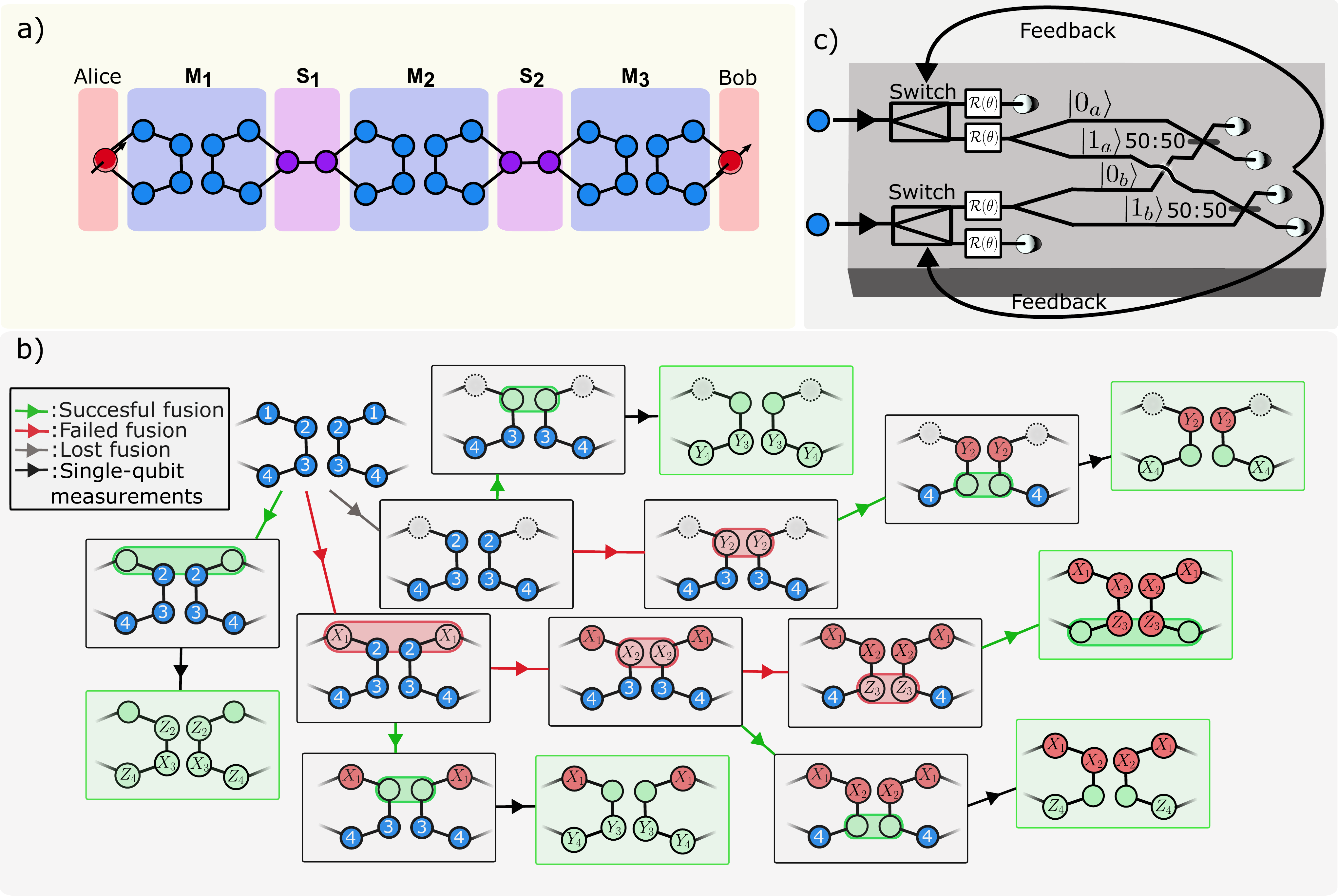}
    \caption{a) Schematics of a quantum repeater setup using four-qubit ring codes. To generate a spin Bell-pair between Alice and Bob, each station ($S_i$) prepares two entangled four-qubit ring states and sends one ring to the left measurement station $M_i$ and one ring to the right station $M_{i+1}$. At the measurement stations, a logical fusion is attempted between the rings, and if logical fusions at all stations succeed, a Bell pair is shared between Alice and Bob. In the figure, blue nodes are photonic qubits, red are stationary spin qubits, and purple nodes are virtual qubits having been measured in the Pauli-$X$ basis. b) Outline of the measurement pattern for a logical fusion between two four-qubit ring codes. The measurement pattern requires no photon delay, as photons are measured in the order they are generated, indicated by the node label. The scheme is adaptive and switches to single qubit measurements once a successful physical fusion has been achieved. The measurement pattern of the single qubit measurements is indicated for every possible successful logical fusion path. Note, however, that not all of these single qubit measurements need to be successful for a successful logical fusion (see Eq.~\eqref{eq:ringfusesucc} for analytical expression of the logical fusion success). c) Illustration of an on-chip photonic circuit to implement the logical fusion measurements. A switch is used to guide the photonic qubits to a fusion measurement or a single qubit measurement, determined by the feedback from the previous measurement outcomes. Before the fusion circuit and the single qubit measurement, Clifford rotations $\mathcal{R}(\theta)$ are applied to adjust the measurement basis.}
    \label{fig:logfusiondecisiontree}
\end{figure*}
In this work, we are mainly interested in ring codes, specifically concatenated ones.
In a ring code, code qubits are entangled in a ring structure~\cite{Tom}, and in a concatenated ring, each code qubit is encoded in another ring.
This is achieved by appending a ring to each code qubit and then measuring the code qubit in the Pauli-$X$ basis~\cite{Codeconcat}.
These code qubits do not need to be explicitly generated since; instead, one can generate the graph state corresponding to the graph state after the Pauli-$X$ measurement transformation~\cite{SinglePauliMeasTransform}. 
The concatenation process can be repeated to create a concatenated ring code with $N$-layers. 
We propose to generate the concatenated rings using memory spin qubits and one spin qubit coupled to a light field.
The latter generates photonic qubits through the sequential application of excitations and spin rotation pulses~\cite{LidnerRudolph,CiracSpinPhotonInterface} (see Fig.~\ref{fig:small_gen_fig} a) and b) for illustration), while the former is used to build successive concatenation layers.
For instance, in Fig.~\ref{fig:small_gen_fig} c), we illustrate the generation of a bare four-qubit ring.
The generation proceeds by entangling the two spins and then emitting four photonic qubits in a line structure.
This is followed by another entangling gate between the spins and a measurement of the optically active spin.
The protocol directly generalizes to more layers, requiring one optically active spin qubit and $N$ memory spin qubits for a $N$-layered concatenated ring, regardless of the size of the unit graph ring.
This is similar to generating a tree with depth $N$~\cite{EconomouRGSGen}.
For a ring code with size $n$~\footnote{For the four-qubit graph code $n=4$} and concatenation depth $N > 1$ (we define $N=1$ for the unit ring), the number of control phase gates between spin qubits needed is found recursively
\begin{equation}
    \label{eq:CZ_gate_count}
    \begin{cases}
    f_{CZ}(N) = n \times f_{CZ}(N-1) + n + 1 \quad \text{if} \quad   N>2 \\
    f_{CZ}(2) = 3n + 1 \quad \text{else},
    \end{cases}
\end{equation}
and the number of spin measurements is
\begin{equation}
    \label{eq:Measurement_count}
    \begin{cases}
    f_{M}(N) = n \times f_{M}(N-1) + 1 \quad \text{if} \quad   N>2 \\
     f_{M}(2) = n + 1 \quad \text{else}.
    \end{cases}
\end{equation}
Furthermore, the number of required photonic qubit emission is
\begin{equation}
    \label{eq:number_of_photonic_qubits_count}
    f_{P}(N) = n^{N}.
\end{equation}
In Appendix~\ref{app:generation_details}, we provide more details on the generation protocol.

\section{Logical fusions with rings}
\begin{figure}
\includegraphics[scale=0.33]{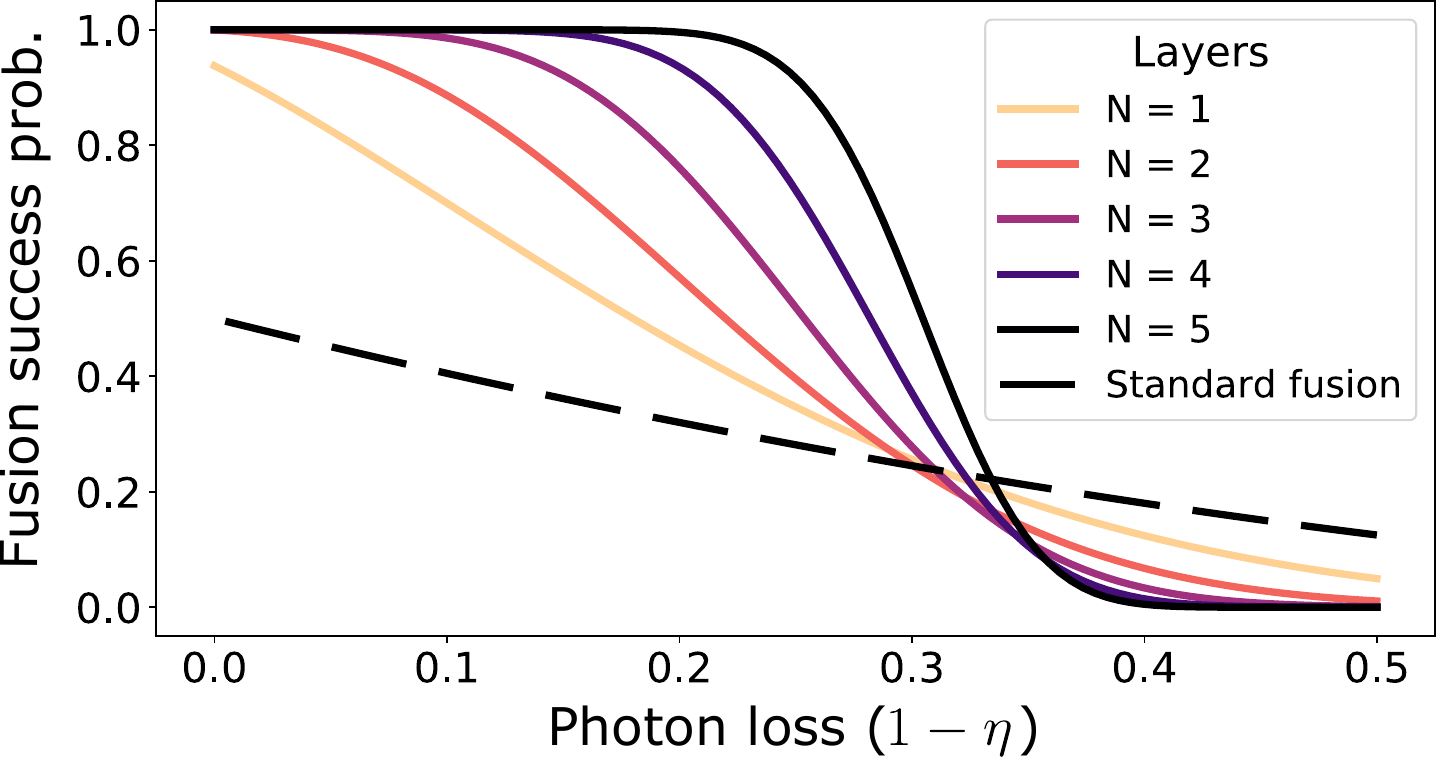}
\caption{The logical fusion success probability as a function of photon loss ($1 - \eta$) for the four qubit rings for concatenation depth up to $N=5$ layers. Here, $N=1$ is the bare ring code shown in Fig.~\ref{fig:logfusiondecisiontree} b). For comparison, the success probability of a standard fusion is illustrated by a dashed line. \label{fig:logfusionsucc}}
\end{figure}

A fusion gate or a Bell state measurement is a probabilistic entangling gate realized with simple linear optical circuits and destructive measurements~\cite{browne2005, Gimeno2016, fusionrules}, illustrated in Fig.~\ref{fig:logfusiondecisiontree} c).
The standard fusion consumes two photonic qubits and, upon success, provides the two parities $X_AX_B$ and $Z_AZ_B$~\footnote{Different Pauli parities can be retrieved upon applying single qubit rotations before the fusion circuit~\cite{fusionrules}} of the two fused photonic qubits $A$ and $B$. 
If the fusion fails, which happens with a probability $p_{\text{fail}} = 50 \%$ for the standard fusion, even in the absence of photon loss, only one of the parities is retrieved, while the other is erased. 
The measured parity upon failure can be chosen by appropriate single-qubit rotations of the photonic qubits before the fusion circuit.
Furthermore, if any of the photonic qubits are lost, both parities are erased.
A way to increase the $50 \%$ success probability of a standard fusion is by introducing $2^{n} - 2$ additional auxiliary photons~\cite{GriceBoosted}, which allows reducing the failure probability to $p_{\text{fail}}=\frac{1}{2^{n}}$.
However, this comes at the cost that none of the $2^n$ photons can be lost. 
An approach to boost the fusion success probability above $50 \%$ while offering protection against photon loss is to encode qubits in graph codes and perform logical fusions~\cite{Tom, pettersson2024, FBQC, AllPhotonicRepeaters}.
Like the standard fusion, a logical fusion between two qubits encoded in a graph code is a measurement pattern of the code qubits that retrieve the two logical parities $\overline{X_AX_B}$ and $\overline{Z_AZ_B}$.
Importantly, because of the redundancy of the graph code, logical fusion can still succeed in case of photon loss of a code qubit.
The measurement pattern to realize a logical fusion can be achieved by either performing physical fusions between all code qubits~\cite{Tom, pettersson2024} or by combining physical fusions and single-qubit measurements~\cite{Tom}.
In the latter case, one can only proceed with single-qubit measurements once a physical fusion between code qubits has succeeded. 
Switching to single-qubit measurements increases the loss-tolerance of the logical fusion as successful single-qubit measurement scales linearly in the transmission efficiency $\eta$ that the photonic qubit experiences, whereas a fusion scales as transmission efficiency squared $\eta^2$.
In this work, we are interested in logical fusions between four-qubit rings since we can devise a particularly simple measurement sequence that does not require reordering photons compared to the order in which they are generated.
The measurement pattern for a logical fusion between two four-qubit ring codes is illustrated in Fig.~\ref{fig:logfusiondecisiontree} b).
We first attempt to fuse the first generated photonic qubits (labeled as 1 in Fig.~\ref{fig:logfusiondecisiontree} b)) with a $XX$ failure basis. 
If the fusion succeeds, we proceed with single qubit measurements, and the logical fusion succeeds if two out of three measurements are successful.
If the fusion failed or a photon was lost, the measurement pattern continues by fusing the second set of generated qubits.
Again, if this fusion succeeds, we proceed with single-qubit measurements; otherwise, we continue fusing code qubits in the order in which they are generated.
Following this procedure, the logical fusion success probability is
\begin{align}
    \label{eq:ringfusesucc}
    \overline{p}_s = p_s(\eta^3 + 3(1-\eta)\eta^2)^2 + p_{f}p_s(\eta^2 + 2(1-\eta)\eta)^2 + \notag\\
    p_lp_s(\eta^4+\eta^2p_{f}) + p_{f}^2p_s(\eta^2 + p_{f}).
\end{align}
Here $p_s = \frac{\eta^2}{2}$ is the probability of a successful physical fusion, $p_{f}=\frac{\eta^2}{2}$ is the probability of fusion failure, and $p_l = 1 - \eta^2$ is the probability that any of the photons are lost in the fusion.
The logical fusion success probability as a function of loss for the four-qubit ring is shown in Fig.~\ref{fig:logfusionsucc} as $N=1$.
As seen in the figure, the logical fusion outperforms standard physical fusion for photon losses up to $30\%$.
To increase the success probability and the loss-tolerance of the logical fusion even further, we can utilize code concatenation, a common approach in quantum error correction~\cite{Codeconcat, Tom}.
Concatenating a four-qubit ring corresponds to encoding each code qubit in another four-qubit ring.
In Fig.~\ref{fig:logfusionsucc}, we plot the logical fusion success probability for up to $N=5$ concatenation layers but leave the details of the logical fusion using concatenated rings for Appendix~\ref{app:concatfusion}.
Already at five layers, we can boost the fusion success probability close to $100 \%$ at $20 \%$ photon loss.
By concatenating further, a $100 \%$ fusion success probability can be sustained at $38 \%$ photon loss. \\

\section{Error correction through error detection}
Although the loss-tolerance capabilities of the concatenated ring are promising, it comes at the cost of an exponential growth of the number of qubits.
In addition to an associated slowdown of communication speed, this also makes it crucial that one can suppress Pauli errors.
Here, we develop a scheme to correct Pauli errors for four-qubit concatenated ring codes for both logical Pauli measurements and logical fusions, which utilize error detection, similar to the scheme studied in Ref.~\cite{ErrorDetectionFT}.
To introduce the concept, we first derive the error correction equations for a logical Pauli measurement and then move on by generalizing it to logical fusions.
\begin{figure}
\includegraphics[scale=0.27]{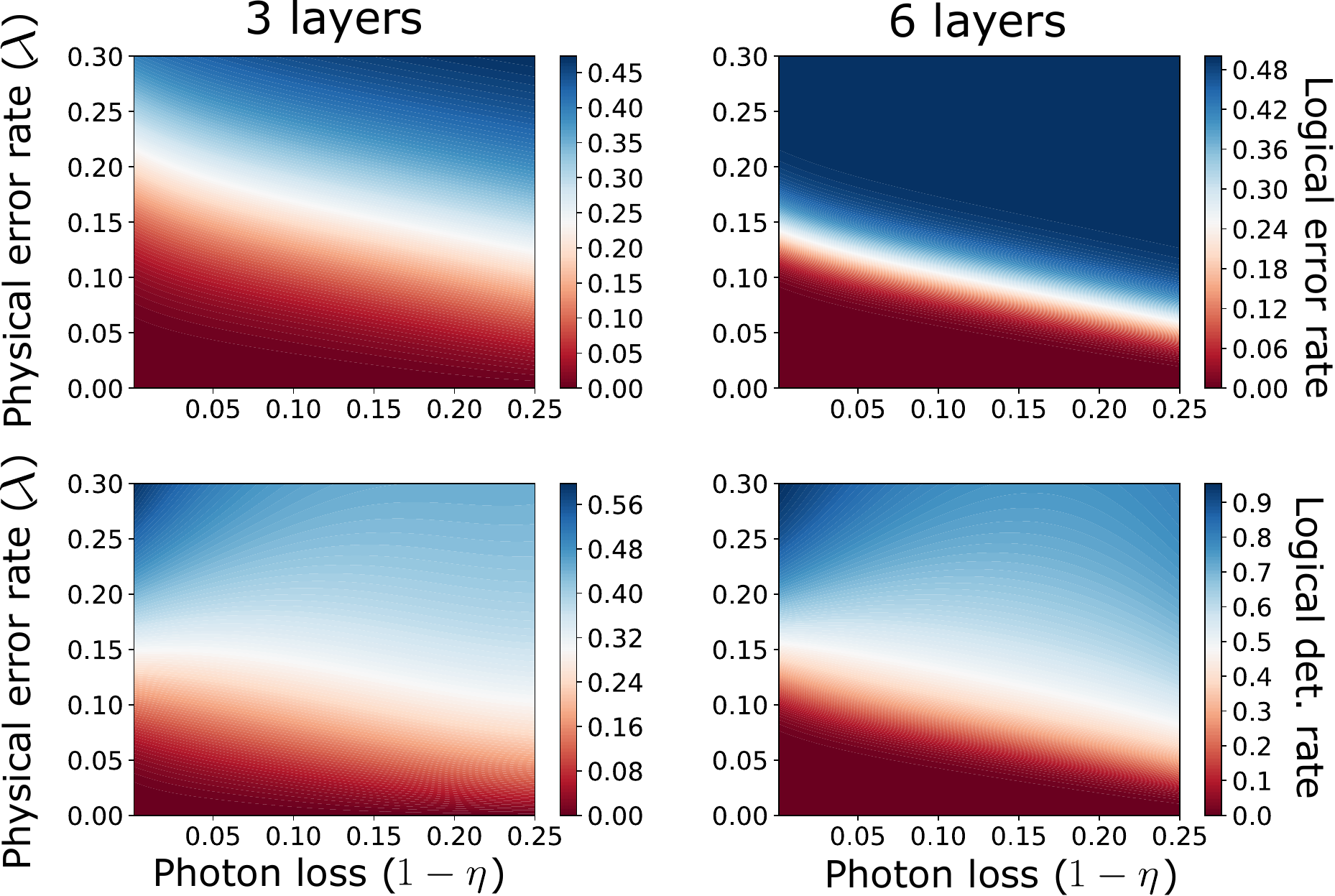}
\caption{The logical error conditioned on not detecting an error (top) and logical detection (bottom) rate of a logical Pauli measurement as a function of photon loss ($1 - \eta$) and physical error rate ($\lambda$) for $N=3$ (left) and $N=6$ (right) concatenation layers. The logical error rate is strongly suppressed for small loss and physical error rates. This suppression increases with the number of layers.\label{fig:pauliMeasurementlossanderror}}
\end{figure}
\subsection{Fault-tolerant logical Pauli measurements}
To perform a logical Pauli measurement of the four-qubit ring code, we consider completely passive measurement patterns for all three Pauli operators. 
Each logical operator can be measured by measuring two different combinations of single qubit operators.
These are
\begin{eqnarray*}
    \overline{X} = \{Z_1\otimes \mathbb{1}_2 \otimes \mathbb{1}_3 \otimes Z_4 , \mathbb{1}_1 \otimes Y_2 \otimes Y_3 \otimes \mathbb{1}_4  \}, \\
    \overline{Y} = \{Y_1\otimes \mathbb{1}_2 \otimes X_3 \otimes \mathbb{1}_4 , \mathbb{1}_1 \otimes X_2 \otimes \mathbb{1}_3 \otimes Y_4  \}, \\
    \overline{Z} = \{X_1 \otimes Z_2  \otimes \mathbb{1}_3 \otimes \mathbb{1}_4 , \mathbb{1}_1 \otimes \mathbb{1}_2 \otimes Z_3 \otimes X_4  \},
\end{eqnarray*}
where the qubit indices are shown in Fig.~\ref{fig:logfusiondecisiontree} b), and $\mathbb{1}$ is the identity operator.
This means that we can measure each of the logical operators by only measuring two out of four photons (e.g., qubits $\{1, 4\}$ or $\{2, 3\}$ for $\overline{X}$).
Thus, the logical transmission is
\begin{equation}
    \label{eq:logical_transmission}
    \overline{\eta} = \eta^4 + 4(1-\eta)\eta^3 + 2(1-\eta)^2\eta^2.
\end{equation}
More importantly, we see that in the case of no loss, the measurement patterns extract two logical operators for each Pauli operator, each with weight two.
This is insufficient to perform error correction~\cite{Tom}, but we can detect errors if the two measured logical operators give disagreeing parities.
This requires a successful measurement of all code qubits, and thus, if the transmission of each photon is taken into account, the detection probability is given by
\begin{equation}
    \label{eq:pauli_detect}
    \epsilon_d = \frac{4\eta^4(\epsilon(1-\epsilon)^3 + \epsilon^3(1-\epsilon))}{\overline{\eta}}
\end{equation}
and the undetected logical error rate is
\begin{equation}
    \label{eq:pauli_error}
    \overline{\epsilon} = \frac{4\eta^4\epsilon^2(1-\epsilon)^2 + (\overline{\eta}-\eta^4)2\epsilon(1-\epsilon)}{\overline{\eta}},
\end{equation}
where $\epsilon$ is the single-qubit measurement error rate.
For simplicity, we here assume each qubit is subjected to a depolarizing channel
\begin{equation}
    \mathcal{E}(\rho) = (1 - \lambda)\rho + \frac{\lambda}{3}\sum_{P \in \{X, Y, Z \}}P\rho P
\end{equation}
where $\rho$ is a single qubit density matrix, and $\lambda$ is the single-qubit error rate.
Following this error model, the single-qubit measurement error rate is related to the single-qubit error rate as $\epsilon = \frac{2 \lambda}{3}$.
We note that this is likely not the correct error model for physical implementations of the scheme since the generation of rings will produce correlated errors~\cite{TomOneEmitterFT}. 
Still, it provides a rough estimate of the performance under faulty operations.
For the bare four-qubit code, in the high-efficiency regime $\eta \sim 1$, we will most likely detect all four photonic qubits, leading to a strong suppression of the logical error.
This shifts the Pauli errors from a logical error to an error detection, which heralds an error but does not allow correcting it.
However, when we concatenate rings, the error detection events from the layers below can be used to correct errors in the layers above.
Error correction in the above layers is achieved by the ability to choose between the two logical operator measurements depending on an error detection event.
For a concatenated ring with $N$-layers, this gives rise to the following recursive equations for the logical error rate and error detection probabilities
\begin{widetext}
\begin{subequations}
\begin{equation}
    \epsilon_{N} = \frac{\eta_{N-1}^4(4\epsilon_{N-1}^2\zeta_{N-1}^2 + 2(4\epsilon_{d, N-1}(1-\epsilon_{d, N-1}) + 2\epsilon_{d, N-1}^2)\epsilon_{N-1}\zeta_{N-1})+ (\eta_N-\eta_{N-1}^4)2\epsilon_{N-1}\zeta_{N-1}}{\eta_N }
\end{equation}
\begin{multline}
\epsilon_{d, N} = ((\eta_N-\eta_{N-1}^4)(2 \epsilon_{d, N-1}(1-\epsilon_{d, N-1}) + \epsilon_{d, N-1}^2) +  
\eta_{N-1}^4(4(\epsilon_{N-1}\zeta_{N-1}^3 + \epsilon_{N-1}^3\zeta_{N-1})  \\
 + 4(\epsilon_{d, N-1}^2(1-\epsilon_{d, N-1})^2 + 
 \epsilon_{d, N-1}^3(1-\epsilon_{d, N-1})) +
 \epsilon_{d, N-1}^4)) 
 /\eta_N ,
\end{multline}
\begin{equation}
    \zeta_{N} = 1 - \epsilon_{N} - \epsilon_{d, N},
\end{equation}
\begin{equation}
    \eta_N = \eta_{N-1}^4 + 4(1-\eta_{N-1})\eta_{N-1}^3 + 2(1-\eta_{N-1})^2\eta_{N-1}^2,
\end{equation}
\end{subequations}
\end{widetext}
where $\zeta_{N}$ is the probability of having neither a detection click nor a Pauli error in layer $N$, and $\eta_0 = \eta$.
In Fig.~\ref{fig:pauliMeasurementlossanderror}, we plot the logical error rate, conditioned on not detecting an error, and the error detection rate as a function of photon loss and physical error rate for a concatenated ring with three and six layers.
For small loss and physical error rates, we see that most error events lead to detection clicks and a resulting strong suppression of the logical error rate. 
As expected, we see that the higher the photon loss $1 -\eta$, the less Pauli errors can be suppressed.
Increasing the concatenation depth, both the logical error rate and detection rate can be suppressed to zero at $\lambda \sim 10\%$ and zero photon loss.

\subsection{Fault-tolerant logical fusion}
\begin{figure}
\includegraphics[scale=0.27]{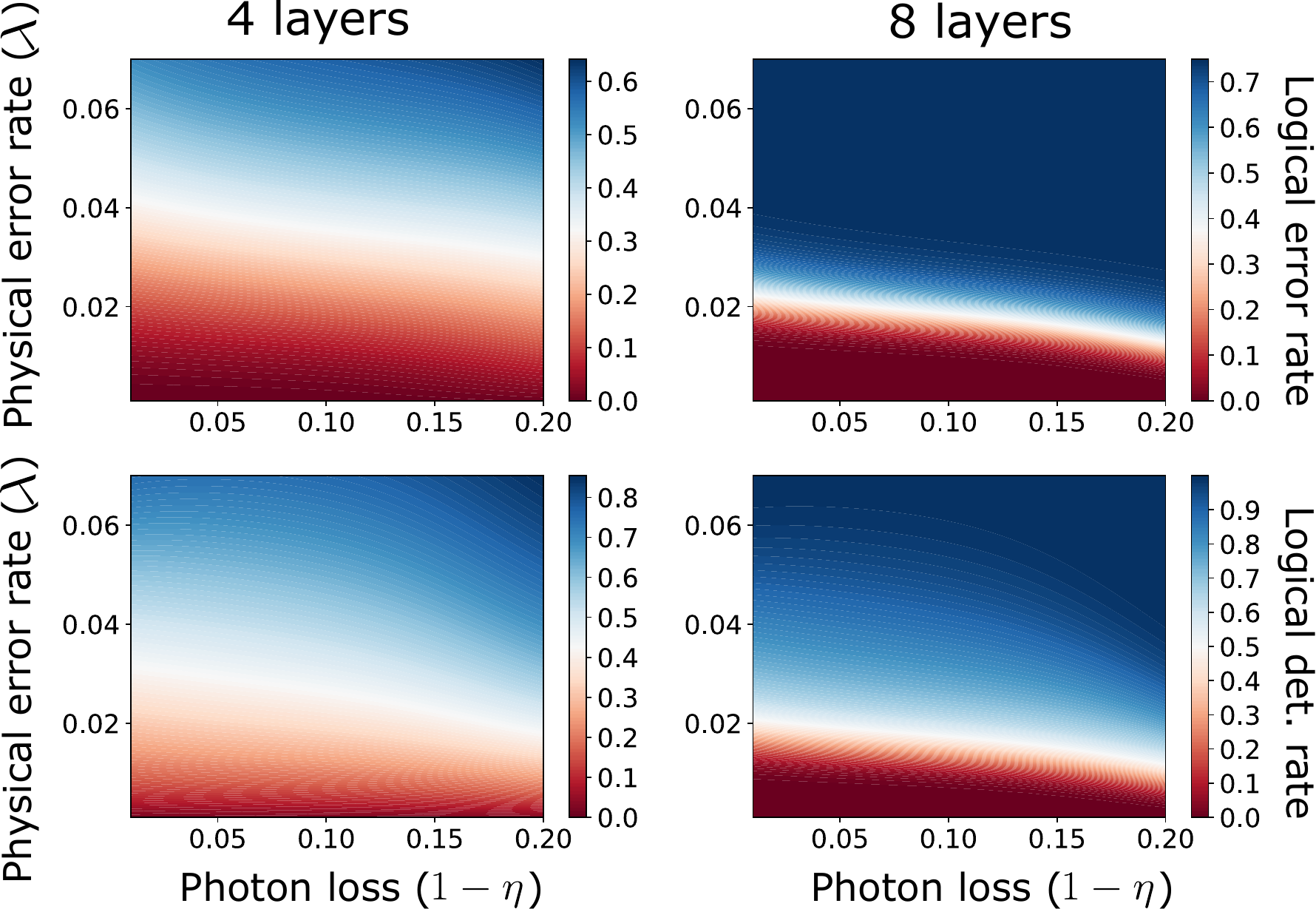}
\caption{The error rate (top) and error detection rate (bottom) of a logical fusion as a function of photon loss ($1 - \eta$) and physical error rate $\lambda$ for $N=4$ (left) and $N=8$ (right) concatenation layers. The logical error rate is strongly suppressed for small loss and physical error rates, with the suppression increasing with concatenation depth. \label{fig:FusionMeasurementlossanderror}}
\end{figure}
As the logical fusion incorporates single-qubit measurements and, thus, logical single-qubit measurements in the upper layers for a concatenated ring, error detection clicks from these measurements can be utilized in the logical fusion.
While this helps to reduce the error rate of the single qubit measurements needed to perform the logical fusion, errors from physical fusions can not be corrected or detected.
This will significantly increase the error rate of the logical fusion.
However, given a concatenated ring with $N$ layers, if one changes the measurement strategy at layer $\Tilde{N}$ to attempt a fusion of all code qubits, several logical parities can be extracted at this layer and above.
With the abundance of parities the same error detection and correction strategy applied for logical Pauli measurements can be utilized for the logical fusion.
Intuitively, this can be understood as having the first $\Tilde{N}$ layers of rings protecting against loss and then additional layers $N - \Tilde{N}$ protecting against errors.
Essential for this scheme is that the loss-tolerant layers provide a fusion success probability $\sim 1$.
In Appendix~\ref{app:fault-tolerant-fusion}, we provide more details on the error correction of logical fusions, and in Fig.~\ref{fig:FusionMeasurementlossanderror}, we plot the logical fusion error and detection rates as a function of photon loss and physical error rate for $N = 4$ and $N = 8$, where we set $\Tilde{N} = 3$.
Here, we plot the logical fusion error conditioned on not detecting an error, and we again assume a depolarizing channel for the single qubit Pauli error rate.
The same behaviour of the logical fusion error and error detection rates is observed as for the logical Pauli measurement in Fig.~\ref{fig:pauliMeasurementlossanderror}.
Furthermore, by increasing the concatenation depth, the logical error and detection rates can essentially be suppressed to zero at $\lambda \sim 1 \%$ and zero photon loss.

\section{Quantum Repeaters}
With the concatenated rings, it is possible to build a quantum repeater setup similar to the setup for repeater graph states (RGS)~\cite{Tom, AllPhotonicRepeaters}.
This setup is illustrated in Fig.~\ref{fig:logfusiondecisiontree} a), where at each station, a two-qubit line is generated, with each line-qubit encoded in a concatenated ring.
Each line qubit is then fused with a line qubit from a neighboring station, and if all fusions succeed, a spin Bell pair is generated between Alice and Bob at the end stations.
Note that to generate the encoded lines, the same number of spins in each emitter is needed as for the concatenated rings, but with double the amount of gate operations.
This is discussed in more detail in Appendix.~\ref{app:generation_details}.
We now analyze the performance of this repeater protocol when subjected to realistic imperfections.

\subsection{Analysis of repeater performance}
The probability of creating a shared spin Bell pair between Alice and Bob is
\begin{equation}
    P_{B} = \overline{p}_s(\eta)^{m+1},
\end{equation}
where $m$ is the number of equally spaced stations, and $\overline{p}_s(\eta)$ is the probability of creating a link between stations (i.e., the logical fusion success probability), which depends on the total efficiency $\eta = \eta_t \eta_d$ each photon experiences.
The total transmission efficiency incorporates out-coupling, frequency conversion, and detection efficiencies in $\eta_d$ and fiber transmission losses in $\eta_t = e^{-L_0/L_{\text{att}}}$.
Here, $L_0$ is the distance between stations and $L_{\text{att}}$ is the attenuation length of an optical fiber at telecom frequencies. 
We use the secret key rate in quantum key distribution to quantify the performance of the repeater protocol.
The effects of operational errors on the secret key rate are included through the secret bit fraction $\mu$ of the transmitted qubits.
Using one-way classical post-processing and perfect classical error correction for the six-state protocol, the secret bit fraction is given by~\cite{SecurityQKD}
\begin{equation}
    \mu(q) = 1- H(q) - q - (1 - q)H(\frac{1 - 3q/2}{1 - q}),
\end{equation}
where $q$ is the error rate of the spin Bell pair shared between Alice and Bob and $H(x) = -x\log_2(x) - (1-x)\log_2(1-x)$ is the binary entropy function.
We assume here that the error probability of the shared Bell pair is $q = 1 - (1 - \epsilon_s)^{m+1}$, where $\epsilon_s$ is the error probability of the logical fusion at each repeater station.
With these quantities, and taking into account the time it takes to generate the graph state at each station $\tau_0$, the secret key rate can be estimated as
\begin{equation}
    \label{eq:standard_rate}
     R = \mu(q) p_s^{m+1} / \tau_0.
\end{equation}
However, as the concatenated rings also have an error detection probability in the logical fusion, we modify the secret key rate for the rings to
\begin{equation}
    \label{eq:ring_rate}
    R_{\text{ring}} = (1-\epsilon_d)^{m+1}\mu(q) p_s^{m+1} / \tau_0,
\end{equation}
where $\epsilon_d$ is the error detection probability of the logical fusion.
If any of the fusions produce an error detection click, the protocol is aborted, which is accounted for by the factor $(1-\epsilon_d)^{m+1}$.

\subsection{Results}
\begin{figure*}[t]
    \centering
    \includegraphics[width=1.0\textwidth]{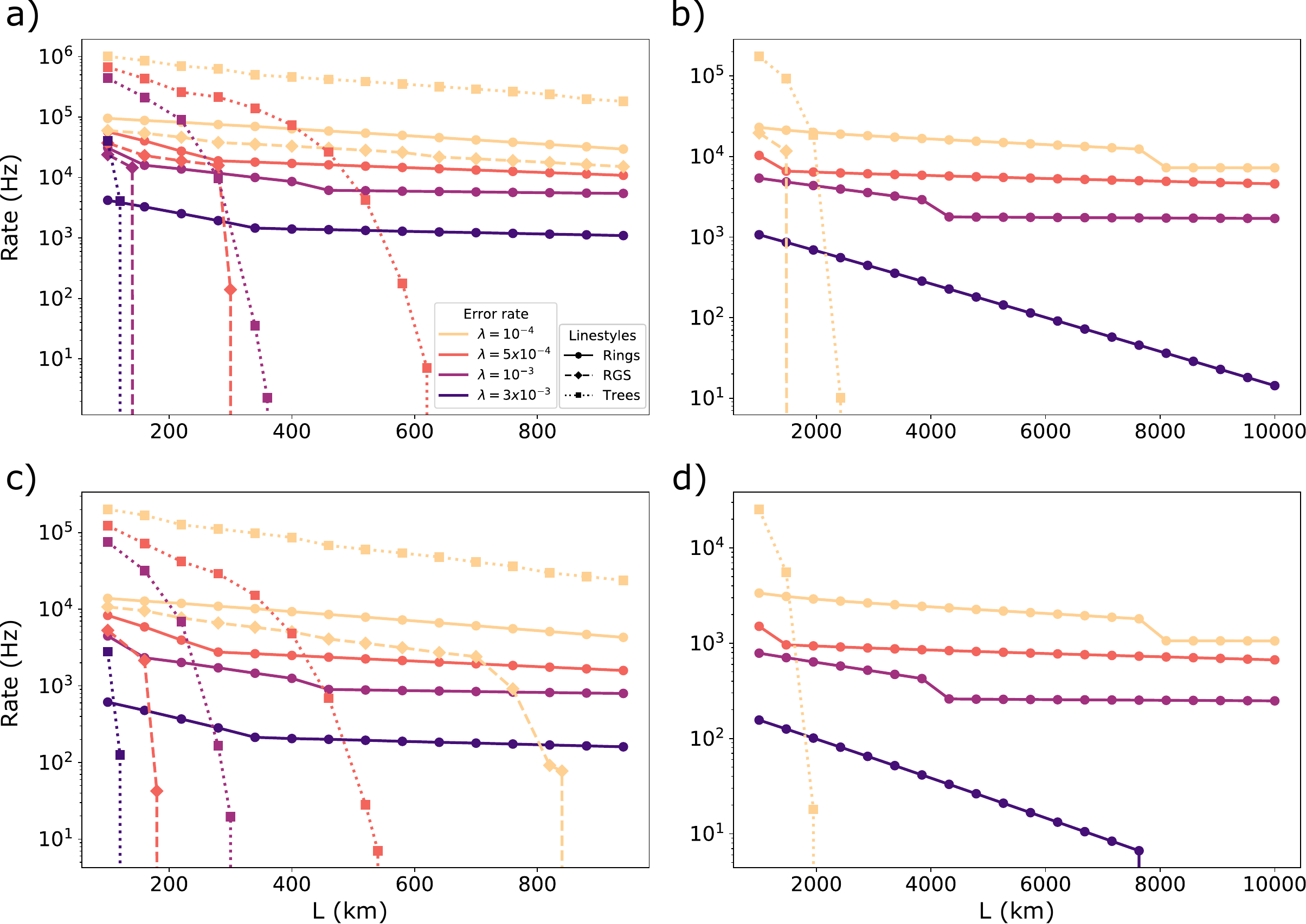}
    \caption{Secret key rate comparison between concatenated rings, RGS~\cite{AllPhotonicRepeaters, SophiaRGS}, and trees~\cite{BorregaardOneWay} for increasing distance ($L$) and various depolarizing error rates ($\lambda$). In all plots, rings are illustrated with solid lines, RGS with dashed lines, and trees with dotted lines. In all four subplots (a-d), we set the photonic qubit generation time to $\tau_{\text{gen}}=1$ ns and the measurement time to $\tau_{M} = 10 \tau_{\text{gen}}$. In a-b), we set the CZ gate time to $\tau_{CZ} = 10$ ns, and in c-d), we set it to $\tau_{CZ} = 100$ ns. a) and c) shows the secret key rates between distances $100 -  1000$ km, whereas b) and d) show the secret key rates between $1000 - 10000$ km.
    }
    \label{fig:rates}
\end{figure*}
To benchmark the repeater performance of the concatenated rings, we compare the achievable secret key rates with the RGS proposal in Ref.~\cite{AllPhotonicRepeaters} and the one-way repeater based on tree graph states~\cite{OriginalTree} proposed in Ref.~\cite{BorregaardOneWay}.
We note that RGS uses probabilistic fusion gates to create entanglement links between stations, like the concatenated rings, while the one-way repeater protocol with trees utilizes spin-photon gates, which are, in principle, deterministic if photon loss is neglected~\cite{ShuoSunSpinPhoton, Reiserer2014}.
We make a comparison with these two repeater protocols since they both require only three matter qubits, similar to the limited number of matter qubits resulting from the linear scaling of the concatenated rings.
However, we emphasize that since the physical requirements for the different schemes are not exactly the same, one should be careful about concluding that one scheme is better than the other only by looking at the resulting rates.
To find the optimal key rates for the three different repeater protocols, we perform a numerical optimization of the cost function
\begin{equation}
    \label{eq:cost_func}
    C = \frac{1}{R}\frac{N_{\text{E}}mL_{\text{att}}}{\tau_{\text{gen}}L},
\end{equation}
where $N_{\text{E}}$ is the required number of matter qubits to generate the graph state, $\tau_{\text{gen}}$ is the photonic qubit emission time, and where $R$ is given in Eq.~\eqref{eq:standard_rate} for the RGS and trees and by Eq.~\eqref{eq:ring_rate} for the concatenated rings.
The cost function seeks to optimize the secret key rate while penalizing the extra cost of adding more repeater stations and more spin qubits~\cite{LukinRepeaterMethodsComparison, BorregaardOneWay}.
Similarly to Ref.~\cite{BorregaardOneWay}, in the optimization, we set a minimum repeater station distance to $L_0 \geq 1$ km, and consider $\eta_d = 0.95$, and $L_{\text{att}} = 20$ km.
Furthermore, we fix $N=7$ as the maximum concatenation layer for the rings.
We note that the RGS and the trees only use three matter qubits, whereas the concatenated rings use $N+1$.
While one can allow for a deeper tree encoding (half of the qubits in the RGS are also encoded in trees~\cite{SophiaRGS, Hilaire2021RGS}), the optimization space becomes intractable if we include this possibility.
Thus, we instead compensate for this by multiplying the achievable secret key rate of the RGS and the trees by an additional factor of $\frac{N+1}{3}$.
In Fig.~\ref{fig:rates}, we plot the optimized secret key rates as a function of the communication length $L$ for the three protocols under various strengths of depolarizing noise. 
This is done for two assumed spin control gate times: in a-b), we assume fast gate times with $\tau_{CZ} = 10$ ns, and in c-d) slower gate times $\tau_{CZ} = 100$ ns.
In both a-b) and c-d), we set the photonic qubit generation time to $\tau_{\text{gen}} = 1$ ns and the measurement time to $\tau_{M} = 10 \tau_{\text{gen}}$, which should be sufficient for a high-fidelity spin measurement given the high efficiency $\eta_d = 0.95$~\cite{SophiaRGS}.
In these figures we see that the trees outperform the rings for small error rates in the fast gate regime. 
This is mainly because the build-up of errors does not affect the rate significantly at short distances, and the significant boost in link creation probability from the spin-photon gate allows for a smaller graph state and, thus, less time spent on generation.
However, as the errors get more significant, the trees suffer from the lack of error correction, and thus, the rings start to outperform them.
In the slow gate regime, the rings are only affected by a lower overall secret key rate, whereas the trees and the RGS are greatly affected by the required delay lines.
These delay lines are required for the trees and RGS as photonic qubits are not measured in the order they are created, which gets longer as the time between photon emission events gets longer.
The effect of the delay lines can be seen from the drop in secret key rate at lower distances $L$ compared to the fast gate times. 
In Appendix~\ref{app:optimazation_details}, we provide more details of the optimizations.
These results are promising for the considered approach, where the concatenated ring can achieve very high rates for extremely long distances ($10^4$ km) in the presence of large single qubit error rates $\epsilon > 10^{-3}$ using very few matter qubits per station ($ \leq 8$).
This presents a resource-efficient repeater protocol to achieve long-distance quantum communication in the presence of significant error rates.

\section{Conclusion and discussion}
We have proposed a one-way quantum repeater protocol based on logical fusions between concatenated rings.
We devised a scheme for generating the concatenated rings using a single quantum emitter coupled to a light field and memory spin qubits, where the number of memory spin qubits scales linearly with the concatenation depth.
The logical fusion necessary for the repeater protocol requires no photon delay lines, as photons are measured in the order in which they are generated. 
Furthermore, we devised a measurement pattern for the logical fusion, which makes it tolerant to both photon loss and Pauli errors already using only five matter qubits, significantly less than previous protocols~\cite{BorregardFaultTolerantOneWay, LukinQPCRepeater100EmitterGenerated}.
Since photon delay lines are not required, the loss tolerance of the logical fusion is robust against slow gate times. 
Furthermore, because of the logical fusion's fault tolerance properties, the achievable secret key rate is resilient against considerable Pauli error rates.
The concatenated rings outperform similar repeater protocols using only a few matter qubits per station and fusion gates for link creation.
Moreover, the concatenated rings even outperform one-way repeaters using deterministic spin-photon gates in the regime of significant error rates or long distances.
For future work, it would be interesting to study more realistic error models of interacting qubits within quantum emitters, as opposed to the depolarizing noise used in this work.
For instance, with the depolarizing noise, making the code bigger allows one to suppress errors better; however, with a more realistic error model considering the spreading of errors from gate operations, errors are likely to become more significant as the graph state gets larger.
Also, it could be interesting to investigate if the adaptive encoding techniques used in Ref.~\cite{GeneralizedRepeater} could be applied to the concatenated rings.
Using these techniques could increase the rate significantly.

\noindent\textit{Acknoweldgements.}
We are grateful to M.C. Löbl, Stefano Paesani, and Oliver August Dall’Alba Sandberg for fruitful discussions. 
We are grateful for financial support from Danmarks Grundforskningsfond (DNRF 139, Hy-Q Center for Hybrid Quantum Networks). L.A.P. acknowledges support from Novo Nordisk Foundation (Challenge project “Solid-Q”). This work was supported by the project QIA-Phase 1, which received funding from the European Union’s Horizon Europe research and innovation program under grant agreement No. 101102140.

\appendix
\section{Generation of concatenated rings}
\label{app:generation_details}
\begin{figure*}[t]
    \centering
    \includegraphics[width=1.0\textwidth]{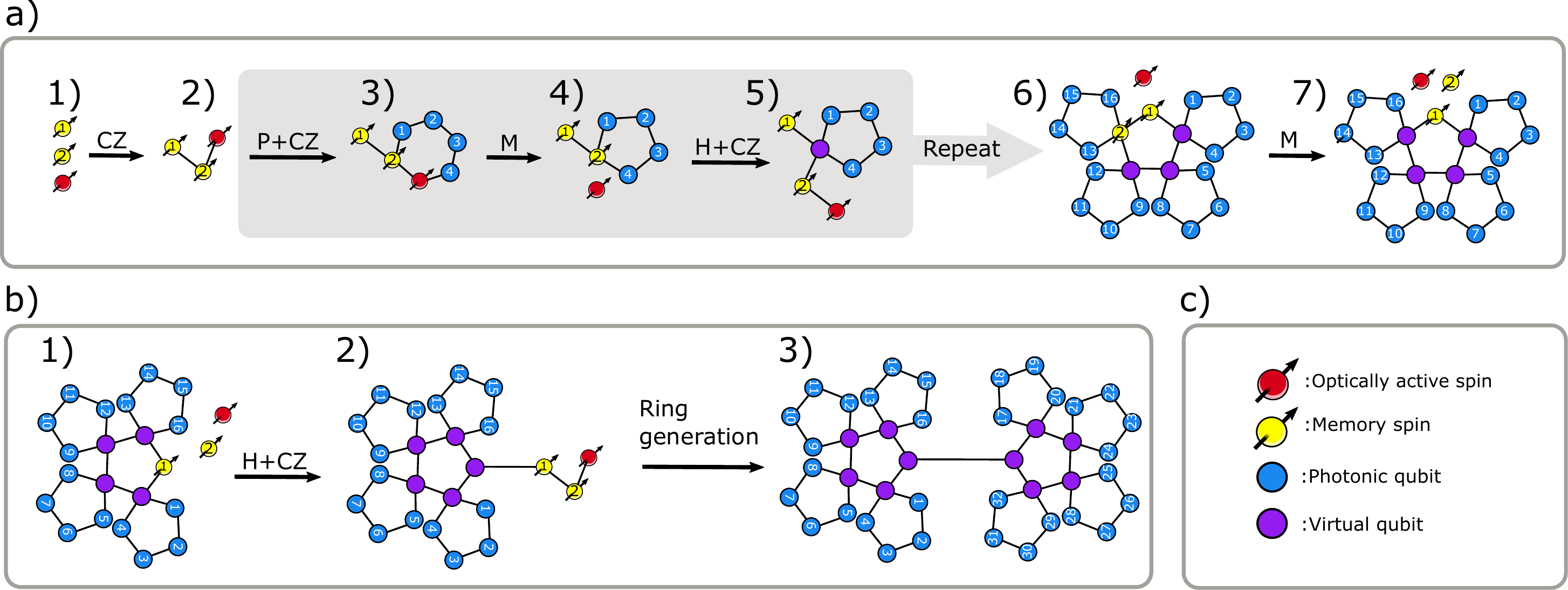}
    \caption{Steps to build a concatenated ring and a two-qubit line concatenated with rings. As defined in c), red nodes symbolize quantum emitters coupled to a light field, yellow nodes memory spin qubits, blue nodes photonic qubits, and purple nodes virtual qubits measured in the Pauli-X basis. Furthermore, $P$ stands for photonic qubit generation, $CZ$ stands for control-Z gates, and $M$ stands for measurements. a) Generation of a two-layered (i.e., N=2) concatenated ring using two memory spin qubits and one spin coupled to a light field. The protocol proceeds through a combination of photonic qubit generation, CZ gates between the spins, and spin measurements, with the full description of the protocol given in the text. b) Generation of a two-qubit line with each qubit encoded in concatenated rings. Step 1) is realized as shown in a). In step 2) a Hadamard gate is applied to memory spin qubit 1, followed by entangling the spins with CZ gates. In step 3) a new concatenated ring is generated as in a) followed by a measurement of memory spin qubit 1 in the Pauli-X basis. }
    \label{fig:concatgen}
\end{figure*}
We propose to generate concatenated rings using memory spin qubits and one spin qubit coupled to a light field as illustrated in Fig.~\ref{fig:concatgen}.
The latter generates spin-photon entanglement through interleaving excitation and spin rotation pulses, as shown in Fig.~\ref{fig:small_gen_fig} a), while the former is used to build successive concatenation layers through control-Z (CZ) gates.
The generation of a concatenated ring with two layers and unit ring size $4$ is sketched in Fig.~\ref{fig:concatgen} a). 
The generation protocol proceeds through the sequence: 1) initialize the spins, 2) entangle all three spins in a line structure, 3) the active spin generates $4$ photonic qubits, 4) a CZ gate between the second memory spin qubit and the optically active spin qubit followed by a Pauli-Y measurement of the optically active spin qubit, 5) perform a Hadamard gate on the second memory spin and entangle it with the optical active spin with a CZ gate, 6) repeat steps 3-5 three more times followed by a CZ gate between the memory spin qubits and a measurement of the second memory spin qubit in the Pauli-Y basis.
As seen in Fig.~\ref{fig:concatgen} a), this generation scheme ensures that the photonic qubits are generated in the order in which they are measured.
The generation of a ring with three concatenation layers is then realized by adding a third memory spin qubit and repeating the process of generating a ring with two concatenation layers.
This generalizes the protocol to $N$ concatenation layers, for which $N$ memory spin qubits and one spin qubit coupled to a light field are needed.
The number of required operations for a concatenated ring with $N$ concatenation layers and $n$ sized unit graph is given in Eqs.~\eqref{eq:CZ_gate_count}-\eqref{eq:number_of_photonic_qubits_count}. \\
The number of required matter qubits is the same to generate two concatenated rings entangled in a two-qubit line structure.
However, the number of operations needed is doubled.
Its generation details are sketched in Fig.~\ref{fig:concatgen} b).
Furthermore, we note that the stations need to generate the left and right line qubits in alternating order to ensure that no delay is required.
\section{Logical fusions with concatenated rings}
\label{app:concatfusion}
In an $N$-layered concatenated ring, each qubit is encoded in a ring itself until the final layers.
Only qubits in the final layer are physical, whereas qubits in the layers above have, in the encoding step, been measured in the Pauli-X basis (either physically or virtually by constructing the corresponding graph code after the measurement). 
The logical fusion success probability for a concatenated ring can be derived from the success probability of the bare ring in Eq.~\eqref{eq:ringfusesucc}.
However, care has to be taken for the logical fusion failure probabilities in the above layers.
The logical fusion success probability, logical fusion failure probability in all three Pauli bases, and logical fusion loss probability of a concatenated ring with $N$-layers are found by recursively updating the equations
\begin{widetext}
    \begin{eqnarray}
        p_{s,N} = p_{s, N-1}(\eta^3_{N-1} + 3(1-\eta_{N-1})\eta_{N-1}^2)^2 + p_{x, N-1}p_{s, N-1}(\eta^2_{N-1} + 2(1-\eta_{N-1})\eta_{N-1})^2 + \nonumber \\
    (p_{l,N-1} + \Tilde{p}_{z,N-1})p_{s, N-1}(\eta^4_{N-1}+\eta^2_{N-1}p_{y, N-1}) + p_{x, N-1}^2p_{s, N-1}(\eta^2_{N-1} + p_{z, N-1}) + \nonumber \\ 
    \Tilde{p}_{x,N-1}p_{l,N-1}p_{s,N-1}\eta^2_{N-1},
    \end{eqnarray}
    \begin{equation}
    \label{eq:ringfusefailX}
    p_{x, N} = p_{l, N-1}p_{s, N-1}(1-\eta_{N-1}^2)(\eta^2_{N-1} + p_{y, N-1}) + p_{x, N-1}^2p_{z, N-1}^2 
    \end{equation}   
\begin{eqnarray}
    \label{eq:ringfusefailZ}
    p_{z, N} = p_{s, N-1}\eta^2_{N-1}\left( 2\eta_{N-1}^2(1- \eta_{N-1})^2 + 4\eta_{N-1} (1- \eta_{N-1})^3 + (1 - \eta_{N-1})^4 \right) \nonumber \\  + p_{x, N-1}p_{s, N-1}(1 - (\eta_{N-1}^2 + 2 \eta_{N-1}(1-\eta_{N-1}))^2) + p_{x, N-1}p_{l, N-1}\eta^4_{N-1} + p_{l, N-1}^2\eta^4_{N-1},
\end{eqnarray}
\begin{equation}
    p_{y, N>1} = 0,
\end{equation}
\begin{equation}
    p_{l, N} = 1 - p_{s, N} - p_{x, N} - p_{y, N} - p_{z, N},
\end{equation}
\end{widetext}
Here, $p_{s, N}$ is the fusion success probability at layer $N$, $p_{i, N}$ is the fusion failure probability in basis $i$ at layer $N$, and $\eta_{N}$ is the logical transmission of a Pauli measurement at layer $N$. 
We denote the last layer as $N=1$, and thus $p_{x, 1} = p_{y, 1} = p_{z, 1} = \frac{\eta^2}{2}$ which is the failure probability of the standard fusion.
Furthermore, the outcomes $\Tilde{p}_{i,N}$ are $\Tilde{p}_{z,N} = p_{z,N}$ and $\Tilde{p}_{x,N} = p_{x,N}$ if $N >1$ and otherwise they are zero.
These represent possible outcomes in the above layers, where a fusion can fail in more than one Pauli basis because they are logical fusions from the layer below.
However, in the final layer, physical fusions are performed where one has to pick one failure basis, and thus these outcomes are not possible (i.e., they are zero).

\section{Fault-tolerant fusion measurement}
\label{app:fault-tolerant-fusion}
To be able to correct the logical parities $\overline{XX}$ and $\overline{ZZ}$, we switch measurement strategy at a certain layer depth $\Tilde{N}$.
This new strategy attempts to fuse all qubits unless a fusion is lost or fails, in which case we switch to single-qubit measurements.
The difference in the new strategy is that we will continue to perform physical fusions even though a successful one has already been obtained.
Adopting this strategy, the logical fusion success probability is
\begin{widetext}
\begin{eqnarray}
    p_{s, N} = p_{s, N-1}^4 +  p_{s, N-1}p_{z, N-1} ((\eta_{N-1}^2 + 2\eta_{N-1}(1 - \eta_{N-1}))^2) \nonumber \\ + p_{s, N-1}p_{x, N-1}\eta_{N-1}^2 +
     p_{s, N-1}(p_{x, N-1} + p_{z, N-1} + 2p_{l, N-1})\eta_{N-1}^4+ \nonumber \\
    p_{s, N-1}^2((p_{l, N-1} + p_{z, N-1}) \eta_{N-1}^2 + p_{x, N-1}) +
    p_{s, N-1}^3(1 - p_{s, N-1}).
\end{eqnarray}
\end{widetext}

In the case we get the $p_{s, N-1}^4$ outcome, two $\overline{XX}$ and $\overline{ZZ}$ parities are retrieved.
If a fusion has been flagged by error detection from the single-qubit measurements in the layers below, or if the parities disagree, we can detect and correct these for the $p_{s, N-1}^4$ outcome where all qubit pairs are successfully fused.
Similarly as for the fault-tolerant Pauli measurement in Eqs.~\eqref{eq:pauli_detect}-\eqref{eq:pauli_error}, the logical error and detection probability for both parities is
\begin{widetext}
\begin{eqnarray}
    \epsilon_{d, N} =p_{s, N-1}^4(4(\epsilon_{N-1}\zeta_{N-1}^3 + \epsilon_{N-1}^3\zeta_{N-1}) + 
  4(\epsilon_{d, N-1}^2(1-\epsilon_{d, N-1})^2 + \nonumber \\
 \epsilon_{d, N-1}^3(1-\epsilon_{d, N-1})) +
 \epsilon_{d, N-1}^4) / p_{s,N}
\end{eqnarray}
\begin{eqnarray}
    \label{eq:fusion_error}
    \epsilon_{N} = p_{s, N-1}^4(4\epsilon_{N-1}^2\zeta_{N-1}^2 + 2(4\epsilon_{d, N-1}(1-\epsilon_{d, N-1}) + 2\epsilon_{d, N-1}^2)\epsilon_{N-1}\zeta_{N-1}) / p_{s, N}
\end{eqnarray}
\begin{equation}
    \zeta_{N} = 1 - \epsilon_{N} - \epsilon_{d, N},
\end{equation}
\end{widetext}
where $\epsilon_{d, N}$ is the error detection rate of the parities at layer $N$, $\epsilon_{N}$ the undetected error rate of the parities at layer $N$, and $\zeta_{N}$ the probability of no error or detection of the parities at layer $N$.
For the other outcomes, no active error correction of the parities can be performed, but errors can still be detected from the Pauli measurements and propagated to the layers above.
Thus, to suppress errors, the fusion success probability at layer $\Tilde{N}$ must be $\sim 1$, such that the error contribution from the non-error correcting outcomes is suppressed.
\section{Optimization details}
\label{app:optimazation_details}
\begin{figure*}[h]
    \centering
    \includegraphics[width=0.95\textwidth]{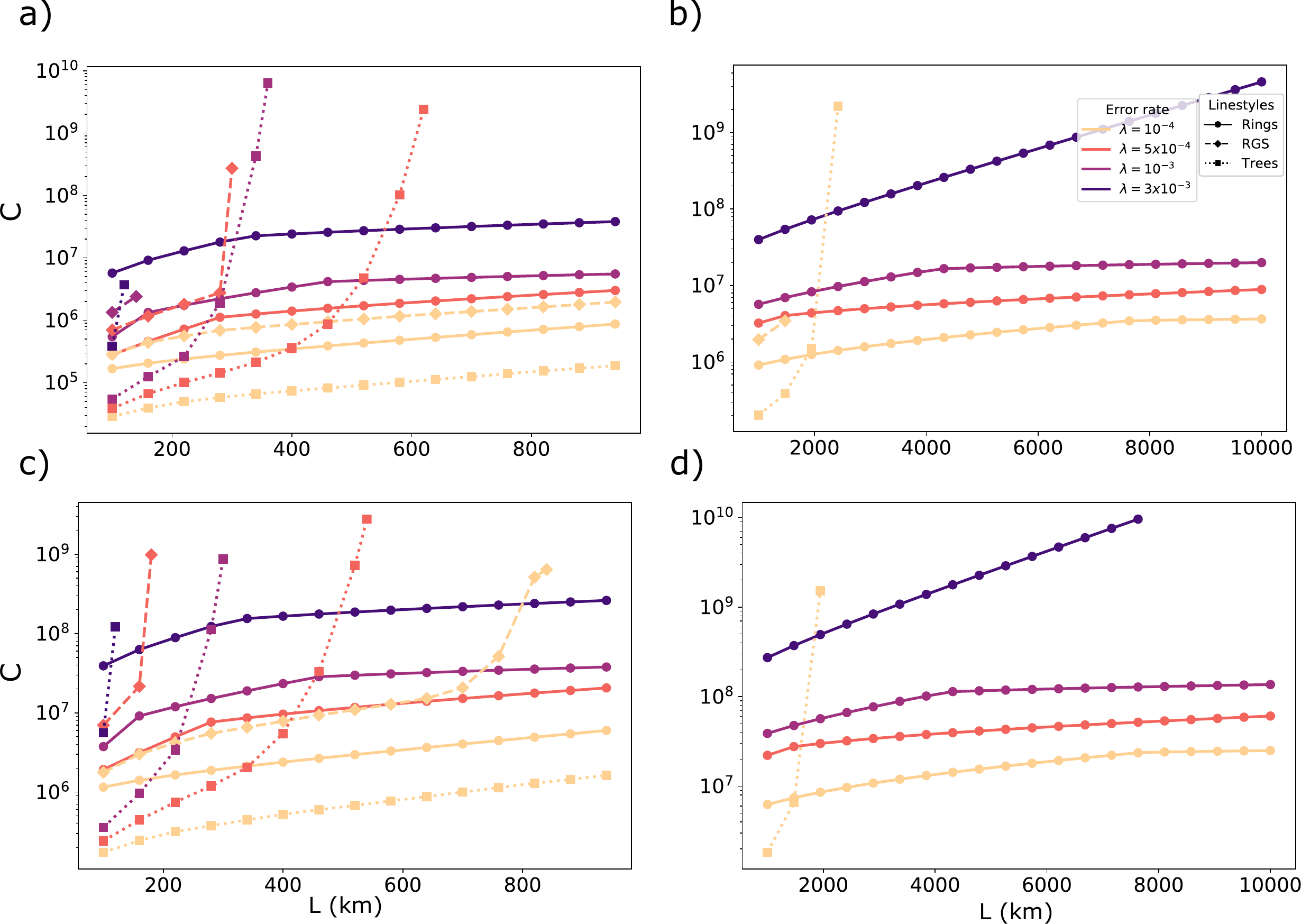}
    \caption{Cost (C) comparison between concatenated rings, RGS~\cite{AllPhotonicRepeaters, SophiaRGS}, and trees~\cite{BorregaardOneWay} for increasing distance ($L$) and various depolarizing error rates ($\lambda$). The parameters in a) to d) are the same as in Fig.~\ref{fig:rates}.}
    \label{fig:costs}
\end{figure*}

\begin{figure*}[h]
    \centering
    \includegraphics[width=0.95\textwidth]{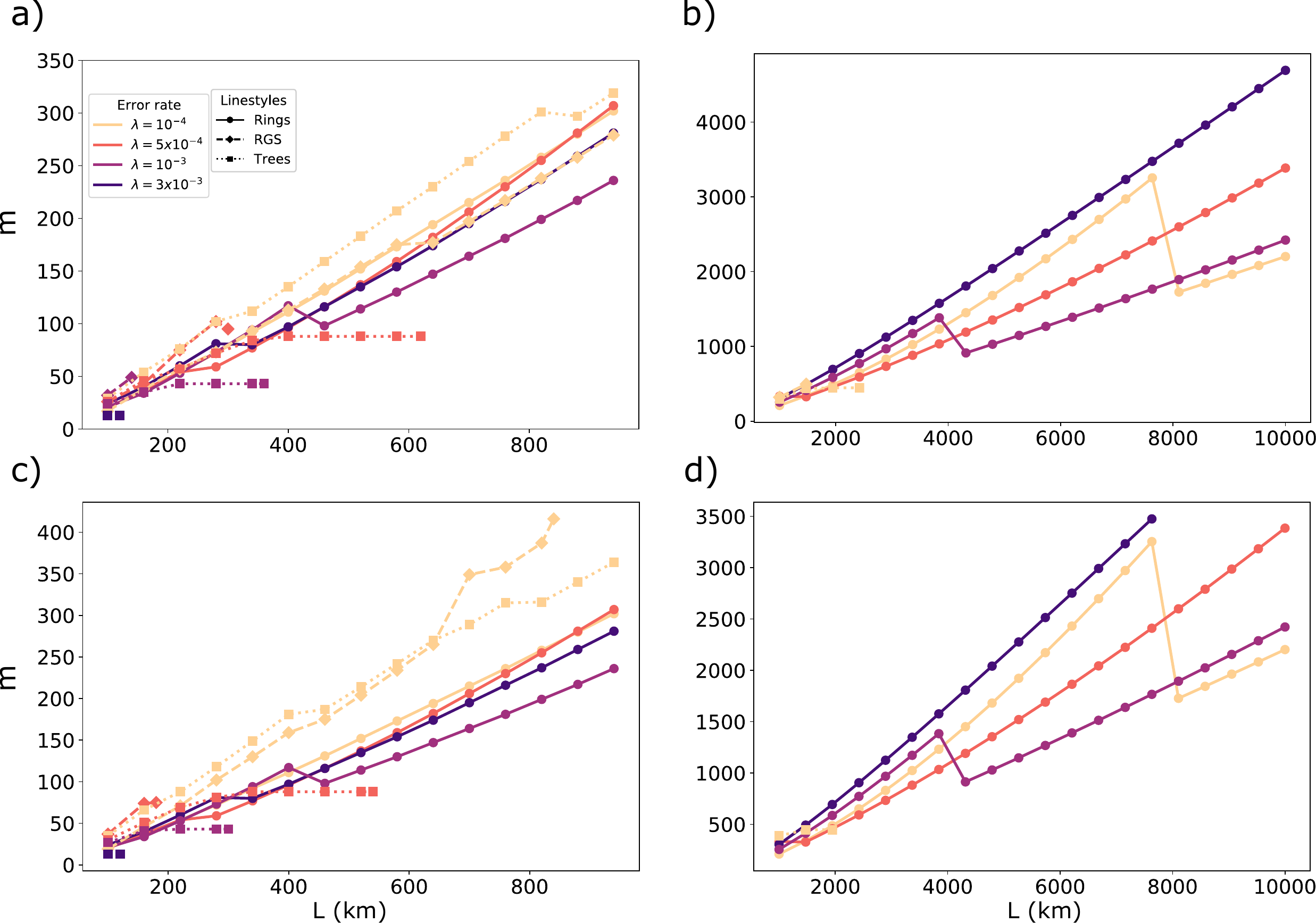}
    \caption{Number of stations (m) comparison between concatenated rings, RGS~\cite{AllPhotonicRepeaters, SophiaRGS}, and trees~\cite{BorregaardOneWay} for increasing distance ($L$) and various depolarizing error rates ($\lambda$). The parameters in a) to d) are the same as in Fig.~\ref{fig:rates}.}
    \label{fig:stations}
\end{figure*}

\begin{figure*}[h]
    \centering
    \includegraphics[width=0.95\textwidth]{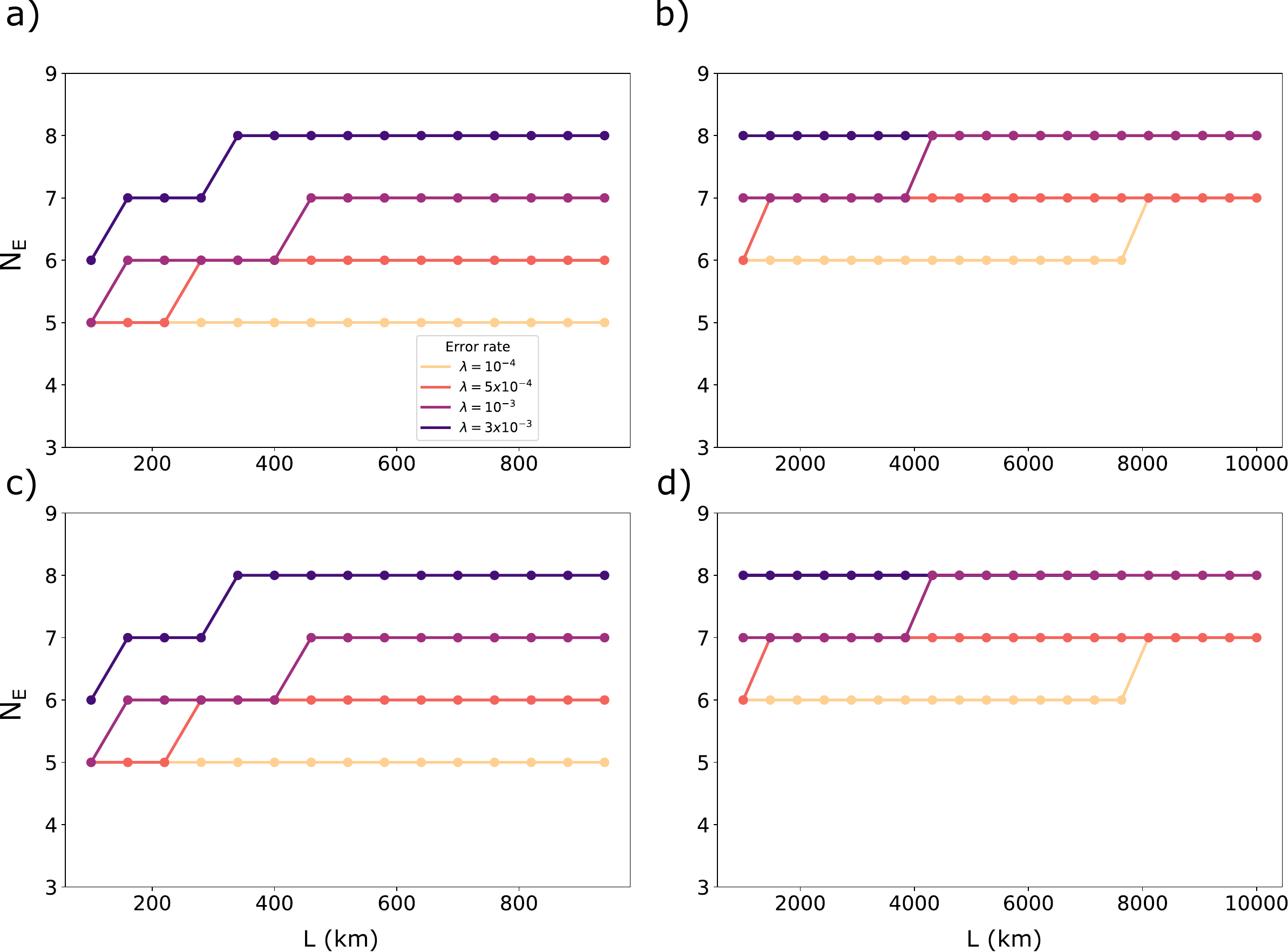}
    \caption{Number of matter qubits ($N_E$) for the concatenated rings for increasing distance and various depolarizing error rates ($\lambda$). The parameters in a) to d) are the same as in Fig.~\ref{fig:rates}. Note that for the trees and the RGS, the number of matter qubits is set to 3. }
    \label{fig:number_of_matter_qubits}
\end{figure*}

This appendix outlines the optimization details in the main text for all three protocols.
Also, we show additional plots of the cost function, the number of repeater stations, and the number of spin qubits to supplement the secret key rate plots shown in Fig.~\ref{fig:rates} in the main text.
For the RGS, we allow up to 40 RGS nodes, keep the tree depth to two, and allow a maximum number of photons of 400 per tree.
For the one-way repeater with trees, we keep the depth limit to three as in Ref.~\cite{BorregaardOneWay} and allow for a maximum size of 2500 qubits. 
Furthermore, for the concatenated rings, we allow for a maximum of seven concatenation layers.
As in Ref.~\cite{BorregaardOneWay}, we enforce a limit to the repeater station spacing of $L_0 \geq 1$ km.
The graph state configurations and number of repeater stations are then picked following an optimization of Eq.~\eqref{eq:cost_func}, which is shown in Fig.~\ref{fig:costs}.
The number of repeater stations is shown in Fig.~\ref{fig:stations}, and the number of matter qubits for the concatenated rings is shown in Fig.~\ref{fig:number_of_matter_qubits}.

\newpage
\bibliography{biblio}

\end{document}